\documentclass[12pt]{article}

\catcode`@=12
\begin{document}
\thispagestyle{empty}
\begin{flushright} 
UCRHEP-T348\\ 
November 2002\
\end{flushright}
\vspace{0.5in}
\begin{center}
{\Large	\bf New and Useful Gauge Extension of the MSSM\\}
\vspace{1.5in}
{\bf Ernest Ma\\}
\vspace{0.2in}
{Physics Department, University of California, Riverside, 
California 92521, USA\\}
\vspace{1.5in}
\end{center}
\begin{abstract}\
A new $nontrivial$ U(1) gauge extension of the Minimal Supersymmetric Standard 
Model (MSSM) is proposed which automatically conserves baryon number and 
lepton number, and solves the $\mu$ problem.  Naturally small $Dirac$ neutrino 
masses are also possible in this context.
\end{abstract}
\vspace{1.0in}

\noindent $\bullet$ Talk at TH-2002, Paris, July 2002.

\newpage
\baselineskip 24pt

If the minimal standard model of quarks and leptons (SM) is extended to 
include supersymmetry, 3 problems are known to appear.  (1) Whereas baryon 
number $B$ and lepton number $L$ are automatically conserved in the SM as 
the consequence of the assumed $SU(3)_C \times SU(2)_L \times U(1)_Y$ 
gauge symmetry and its representation content, they are no longer so in its 
most general supersymmetric extension.  (2) The necessity of 2 Higgs doublet 
superfields requires the term $\mu \hat \phi_1 \hat \phi_2$, but there is 
no understanding of why $\mu$ should be of the order of the supersymmetry 
breaking scale $M_{SUSY}$ instead of some much larger unification scale. 
(3) Neutrino masses are not required by the gauge symmetry, which is of 
course also a shortcoming of the SM.

Problem (1) is usually solved by imposing $R$ parity, i.e. $R \equiv 
(-1)^{2j+3B+L}$, which is of course the defining hypothesis of the Minimal 
Supersymmetric Standard Model (MSSM).  Problem (2) is usually ignored or 
solved by replacing $\mu$ by a singlet superfield $\hat S$ with 
$\langle \tilde S \rangle \sim M_{SUSY}$.  Problem (3) is usually sidestepped 
as in the SM by adding trivial $\hat N^c$ singlets.

Each of the above solutions is independent of the other two.  Is there a 
single principle which solves all 3 problems?  The answer is yes, as I 
show in this talk, using a new $nontrivial$ U(1) gauge symmetry, as recently 
proposed \cite{ma02}.

Consider the gauge group $SU(3)_C \times SU(2)_L \times U(1)_Y \times U(1)_X$. 
The usual quark and lepton (left-handed) chiral superfields transform as 
follows:
\begin{eqnarray}
&& (\hat u, \hat d) \sim (3,2,1/6;n_1), ~~ \hat u^c \sim (3^*,1,-2/3;n_2), 
~~ \hat d^c \sim (3^*,1,1/3;n_3), \\ 
&& (\hat \nu, \hat e) \sim (1,2,-1/2;n_4), ~~ \hat e^c \sim (1,1,1;n_5), 
~~ \hat N^c \sim (1,1,0;n_6).
\end{eqnarray}
They are supplemented by the two Higgs doublet superfields
\begin{equation}
\hat \phi_1 \sim (1,2,-1/2;-n_1-n_3), ~~~ \hat \phi_2 \sim (1,2,1/2;-n_1-n_2),
\end{equation}
with $n_1 + n_3 = n_4 + n_5$, and $n_1 + n_2 = n_4 + n_6$ to allow for the 
usual Yukawa couplings of the quarks and leptons as in the MSSM.  However, 
the $\mu$ term is replaced by the trilinear interaction $\hat \chi \hat 
\phi_1 \hat \phi_2$, where $\hat \chi$ is a Higgs singlet superfield 
transforming as
\begin{equation}
\hat \chi \sim (1,1,0;2n_1+n_2+n_3).
\end{equation}
Thus $2n_1 + n_2 + n_3 \neq 0$ is required so that the effective $\mu$ 
parameter of this model is determined by the $U(1)_X$ breaking scale, i.e. 
$\langle \hat \chi \rangle$.  Furthermore, $n_4 \neq -n_1-n_3$ would imply 
$L$ conservation, and $n_2+2n_3 \neq 0$ would imply $B$ conservation. 
In contrast to the SM, the existence of $\hat N^c$ is required here 
as long as $n_6 \neq 0$.

The big question is now whether there is an anomaly-free $U(1)_X$ gauge 
symmetry with the above properties.  The answer is yes if I add two copies 
of the singlet quark superfields
\begin{equation}
\hat U \sim (3,1,2/3;n_7), ~~~ \hat U^c \sim (3^*,1,-2/3;n_8),
\end{equation}
and one copy of
\begin{equation}
\hat D \sim (3,1,-1/3;n_7), ~~~ \hat D^c \sim (3^*,1,1/3;n_8),
\end{equation}
with $n_7 + n_8 = -2n_1 - n_2 - n_3$, so that their masses are also 
determined by the $U(1)_X$ breaking scale.

Consider now the various conditions for the absence of the axial-vector 
anomaly \cite{ava}.  The $[SU(3)]^2 U(1)_X$ anomaly is absent automatically 
because $2n_1 + n_2 + n_3 + n_7 + n_8 = 0$.  The $[SU(2)]^2 U(1)_X$ and 
$[U(1)_Y]^2 U(1)_X$ anomalies are both absent if $n_2+n_3 = 7n_1+3n_4$. 
The $U(1)_Y [U(1)_X]^2$ anomaly is the sum of 8 quadratic terms in $n_i$, 
but it factorizes remarkably into
\begin{equation}
6(3n_1+n_4)(2n_1-4n_2-3n_7) = 0.
\end{equation}
The condition $3n_1+n_4 = 0$ contradicts $2n_1+n_2+n_3 \neq 0$, so the 
condition $2n_1 - 4n_2 - 3n_7 = 0$ has to be chosen.

The $[U(1)_X]^3$ anomaly is the sum of 11 cubic terms in $n_i$, but it 
factorizes even more remarkably into
\begin{equation}
-36 (3n_1+n_4) (9n_1+n_4-2n_6) (6n_1-n_4-n_6) = 0.
\end{equation}
Thus there are 2 solutions, as summarized in Tables 1 and 2.

\begin{table}[htb]
\caption{Solutions (A) and (B) where $n_i = a n_1 + b n_4$.}
\begin{center}
\begin{tabular}{||c|c|c||c|c||}
\hline \hline
& \multicolumn{2}{c||}{(A)} & \multicolumn{2}{c||}{(B)} \\ 
\cline{2-5}
 & $a$ & $b$ & $a$ & $b$ \\ 
\hline
$n_2$ & 7/2 & 3/2 & 5 & 0 \\ 
$n_3$ & 7/2 & 3/2 & 2 & 3 \\ 
$n_5$ & 9/2 & 1/2 & 3 & 2 \\ 
$n_6$ & 9/2 & 1/2 & 6 & --1 \\ 
$n_7$ & --4 & --2 & --6 & 0 \\ 
$n_8$ & --5 & --1 & --3 & --3 \\ 
\hline
$-n_1-n_3$ & --9/2 & --3/2 & --3 & --3 \\ 
$-n_1-n_2$ & --9/2 & --3/2 & --6 & 0 \\ 
$2n_1+n_2+n_3$ & 9 & 3 & 9 & 3 \\ 
\hline \hline
\end{tabular}
\end{center}
\end{table}

\begin{table}[htb]
\caption{Conditions on $n_1$ and $n_4$ in (A) and (B).}
\begin{center}
\begin{tabular}{||c|c||c|c|l||}
\hline \hline
\multicolumn{2}{||c||}{(A)} & \multicolumn{2} {c|}{(B)} & \\ 
\hline
$c$ & $d$ & $c$ & $d$ & $cn_1+dn_4 \neq 0$ forbids \\ 
\hline
3 & 1 & 3 & 1 & $\mu$ term \\ 
9 & 5 & 3 & 4 & $L$ violation \\ 
7 & 3 & 3 & 2 & $B$ violation \\ 
1 & 1 & 1 & 3 & $U^c$ as diquark \\ 
1 & 1 & 1 & 0 & $D^c$ as diquark \\ 
1 & 0 & 5 & --1 & $U$ as leptoquark \\ 
1 & 0 & 1 & 1 & $D$ as leptoquark \\ 
13 & 1 & 4 & 3 & $U^c$, $D^c$ as semiquarks \\ 
\hline \hline
\end{tabular}
\end{center}
\end{table}

Note that solutions (A) and (B) are identical if $n_4 = n_1$.  This turns 
out to be also the condition \cite{coc} for $U(1)_X$ to be orthogonal to 
$U(1)_Y$.  The condition $3n_1+n_4 \neq 0$ also forbids $\hat Q \hat Q 
\hat Q \hat L$ and $\hat u^c \hat u^c \hat d^c \hat e^c$ which are allowed 
by $R$ parity.  Thus proton decay is more suppressed here than in the MSSM.

There are two more anomalies to consider.  The global SU(2) chiral gauge 
anomaly \cite{witten} is absent because the number of $SU(2)_L$ doublets 
is even.  The mixed gravitational-gauge anomaly \cite{mixed} is proportional 
to the sum of $U(1)_X$ charges, i.e.
\begin{eqnarray}
&& 3(6n_1 + 3n_2 + 3n_3 + 2n_4 + n_5 + n_6) + 3(3n_7 + 3n_8) \nonumber \\ 
&& + 2(-n_1-n_3) + 2(-n_1-n_2) + (2n_1 + n_2 + n_3) = 6(3n_1 + n_4),
\end{eqnarray}
which is not zero.  This anomaly may be tolerated if gravity is neglected. 
On the other hand, it may be rendered zero by adding $U(1)_X$ 
supermultiplets as follows: one with charge $3(3n_1+n_4)$, three with charge 
$-2(3n_1+n_4)$, and three with charge $-(3n_1+n_4)$.  Hence they contribute 
$3 + 3(-2-1) = -6$ (in units of $3n_1+n_4$) to Eq.~(9), but $27 + 3(-8-1) 
= 0$ to Eq.~(8).

As it stands, this model pairs $\nu$ with $N^c$ to form a Dirac neutrino 
with mass proportional to $\langle \phi_2 \rangle$.  This would require 
extremely small Yukawa couplings and is generally considered to be very 
unnatural.  On the other hand, if $n_6 = 3n_1 + n_4$ is assumed [i.e. 
$n_4 = 3n_1$ in solution (A) or $n_4 = 3n_1/2$ in solution (B)], then the 
number and structure of the extra singlets used to cancel the mixed 
gravitational-gauge anomaly of Eq.~(9) are exactly right to allow neutrinos 
to acquire naturally small seesaw Dirac masses, as shown below.

In addition to the 3 singlets $N^c$ of $U(1)_X$ charge $n_6$, there are now 
also 3 singlets $N$ of charge $-n_6$ and 3 singlets $S^c$ of charge $-2n_6$. 
The $12 \times 12$ mass matrix spanning $(\nu, S^c, N, N^c)$ is then of the 
form
\begin{equation}
{\cal M} = \left[ \begin{array} {c@{\quad}c@{\quad}c@{\quad}c} 0 & 0 & 0 & m_1 
\\ 0 & 0 & m_2 & 0 \\ 0 & m_2 & 0 & M \\ m_1 & 0 & M & 0 \end{array} \right],
\end{equation}
where $m_1$ comes from $\nu N^c \phi_2^0$ with $\langle \phi_2^0 \rangle 
\neq 0$, $m_2$ comes from $N S^c \chi$ with $\langle \chi \rangle \neq 0$, 
and $M$ is an allowed invaraint mass.  Thus $m_1 \sim 10^2$ GeV, $m_2 \sim 
10^3$ GeV, and $M \sim 10^{16}$ GeV are expected.  In the reduced $(\nu, S^c)$ 
sector, the effective $6 \times 6$ mass matrix is still exactly of the Dirac 
form, i.e.
\begin{equation}
{\cal M}_\nu = \left[ \begin{array} {c@{\quad}c} 0 & -m_1 m_2/M \\ -m_1 m_2/M 
& 0 \end{array} \right],
\end{equation}
and $m_1 m_2/M \sim 10^{-2}$ eV is the right order of magnitude for 
realistic neutrino masses.

In conclusion, a remarkable new U(1) gauge symmetry has been identified in a 
simple extension of the supersymmetric standard model which is capable of 
enforcing $B$ or $L$ conservation or both, as well as the absence of the $\mu$ 
term and the presence of naturally small Dirac neutrino masses. The origin of 
this new U(1) gauge symmetry is unknown at present; it has no obvious fit 
into any simple model of grand unification or string theory.

This work was supported in part by the U.~S.~Department of Energy under 
Grant No.~DE-FG03-94ER40837.

%\newpage
\bibliographystyle{unsrt}

\end{document}